# Encapsulation of phosphorus dopants in silicon for the fabrication of a quantum computer

L. Oberbeck, N. J. Curson, M. Y. Simmons, R. Brenner, A. R. Hamilton, S. R. Schofield, and R. G. Clark

*Australian Research Council Special Research Centre for Quantum Computer Technology, School of Physics, The University of New South Wales, Sydney NSW 2052, Australia*

The incorporation of phosphorus in silicon is studied by analyzing phosphorus δ-doped layers using a combination of scanning tunneling microscopy, secondary ion mass spectrometry and Hall effect measurements. The samples are prepared by phosphine saturation dosing of a Si(100) surface at room temperature, a critical annealing step to incorporate phosphorus atoms, and subsequent epitaxial silicon overgrowth. We observe minimal dopant segregation (~5 nm), complete electrical activation at a silicon growth temperature of 250 °C and a high two-dimensional electron mobility of ~$10^2$ cm$^2$/Vs at a temperature of 4.2 K. These results, along with preliminary studies aimed at further minimizing dopant diffusion, bode well for the fabrication of atomically precise dopant arrays in silicon such as those found in recent solid-state quantum computer architectures.

From Moore's Law it is well-known that the number of transistors on a chip doubles approximately every 18 months.[1] In order to maintain this trend alternative means of fabricating devices are actively being pursued[2] and it is clear that the ability to fabricate atomically precise structures in silicon is becoming increasingly important. It is also important to characterize the spatial and electrical distribution of the dopant atoms. This is especially the case for a number of proposals to fabricate atomically precise dopant arrays in silicon for the fabrication of silicon based quantum computers.[3-5] In these architectures the dopant atom is required to be electrically active. The free electron of each dopant can then either act directly as the quantum bit[5] or mediate the coupling between quantum bits.[3,4]

Recent results have demonstrated that a precise array of phosphorus bearing molecules can be fabricated on a silicon surface using a hydrogen resist based strategy.[6] The next important step for creating ordered dopant arrays in silicon devices, which has not yet been demonstrated, is to encapsulate the dopants in high quality epitaxial silicon without disturbing the array. This step must aim at choosing an optimal substrate temperature to minimize dopant diffusion and surface segregation during growth, while maintaining a high structural quality of the epitaxial layer.

While numerous publications exist on B and Sb δ-doping in Si,[7,8] P δ-doping has been applied only recently to the fabrication of SiGe tunneling diodes.[9,10] These devices are of great interest for digital and high frequency applications due to their negative differential resistance. However, high peak to valley current ratios of about 5 have to be achieved for realistic applications, which requires minimal diffusion of the dopants in the device.[9] This has only recently been demonstrated in the fabrication of SiGe tunneling diodes, where P δ-doped layers are fabricated with a GaP solid dopant source.[9,10]

This Letter describes recent progress in the low temperature encapsulation of phosphorus dopants in silicon and represents one of the first demonstrations of P δ-doping using phosphine gas as the dopant source.[11] The use of phosphine (PH$_3$) gas has previously been shown – with the deposition of a hydrogen resist layer and atomic lithography using a scanning tunneling microscopy (STM) tip – to allow atomic precision doping of the Si surface.[6] We now present STM, secondary ion mass spectrometry (SIMS) and Hall resistance data of P δ-doped layers in Si to determine whether phosphine doping is compatible with complete electrical activation of the dopants with minimal diffusion during encapsulation. STM results show that a critical annealing step allows incorporation of the phosphorus donors in substitutional lattice sites before low temperature silicon encapsulation. Subsequent Hall measurements show that all of the donors are electrically active, and SIMS analysis displays minimal spreading of the donors after low temperature overgrowth (<5 nm). These results represent significant progress towards the fabrication of atomically precise dopant profiles in silicon and in particular for the fabrication of a solid-state quantum computer using P dopants in Si.

Figure 1 (a) to (e) shows STM images of the fabrication steps used to prepare P δ-doped layers in Si. Clean Si(100) surfaces are produced by flashing samples of resistance ~1 Ωcm for 30 s at a temperature of ~1150 °C (Fig. 1 (a)). The sample is then saturation dosed with PH$_3$ and the corresponding STM image (Fig. 1 (b), (c)) clearly shows local regions of c(4×2) ordering of the adsorbed phosphine molecules on the silicon surface.[12] A critical stage of the fabrication process is then to anneal the sample for 5 min at 550 °C, which not only incorporates P atoms into the surface with a density of 0.25 monolayers (ML)[13] (corresponding to a concentration of $1.7 \times 10^{14}$ cm$^{-2}$) but also desorbs the hydrogen[14] (Fig. 1 (d)). This sheet of P atoms is then overgrown by a 24 nm thick epitaxial i-Si film (ρ > 1 kΩcm) deposited at a substrate temperature of 250 °C (Fig. 1 (e)).

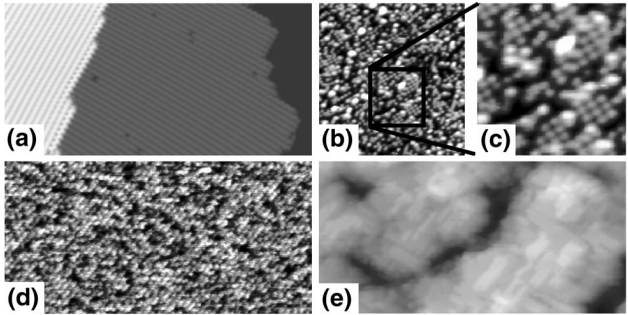

FIG. 1. STM images of the fabrication process for a P δ-doped layer in Si: (a) clean Si(100) at -1.6V, (b), (c) PH$_3$ saturation dosed Si surface at –3V, (d) saturation dosed surface after critical annealing step, and (e) epitaxial Si overgrowth of the saturation dosed surface at –1.4V. Images display an area of $50 \times 25$ nm$^2$ except for (b) $25 \times 25$ and (c) $10 \times 10$ nm$^2$.

The significance of the critical anneal step is demonstrated further in Fig. 2. In Fig. 2 (a), (c) we lightly dose a clean Si(100) surface at room temperature with PH$_3$. Bright features are observed in the STM images with a height profile of ~0.15 nm (Fig. 2 (e)) above the Si surface corresponding to the adsorption of PH$_3$, which is known to quickly dissociate to PH$_2$.[13] After annealing at ~ 530 °C the P atoms are incorporated into the Si surface to form Si-P heterodimers with a reduced height of ~0.04 nm above the surface, shown in Fig. 2 (b), (d), (f). This STM investigation shows for the first time direct evidence of the incorporation of P atoms into silicon to form Si-P heterodimers, from adsorbed phosphine molecules, after a critical annealing step. The annealing step not only enhances the electrical activation of the P atoms but helps to minimize segregation during subsequent Si growth since they are now bound to the Si surface with three, rather than one, covalent bonds. A detailed STM/Auger spectroscopy study of this incorporation process will be published elsewhere.

SIMS measurements were performed on the δ-doped samples to determine the level of surface segregation and diffusion of the P atoms during epitaxial silicon overgrowth. Using an ATOMIKA 6500 system with Cs$^+$ primary ion beam the P depth profile was recorded at energies of 2 (not shown here), 5.5 and 15 keV (Fig. 3 (a)). Two distinct mass-31 peaks at a depth of ~5 and ~24 nm are recorded. Separate measurements using a CAMECA SIMS system with a higher mass resolution confirmed that the peak at 5 nm is due to $^{30}$SiH arising from H adsorption at the Si surface and not to $^{31}$P. The peak at 24 nm however was ascribed to the P δ-doped layer at the interface between the epitaxial layer and the substrate. This peak is observed to decay into the substrate with the decay length increasing from ~1 decade/10 nm to ~1 decade/20 nm as the primary ion energy is increased. This variation of the decay length with ion energy is a well-known measurement artefact arising from ion beam mixing, i.e. primary Cs$^+$ ions pushing P atoms further into the substrate during the sputtering process. The SIMS data show a full width at half maximum of the $^{31}$P peak of 5, 6, and 14 nm for Cs$^+$ primary ion energies of 2, 5.5, and 15 keV, respectively. These results demonstrate that the P atoms have moved vertically only a few nm from their initial positions during the silicon overgrowth step at 250°C. Additionally it is possible to determine the total concentration of P atoms in the δ-doped layer from the SIMS signal, which was found to be $1.4 \pm 0.3 \times 10^{14}$ cm$^{-2}$, in close agreement with the concentration expected for 0.25 ML of P atoms of $1.7 \times 10^{14}$ cm$^{-2}$.

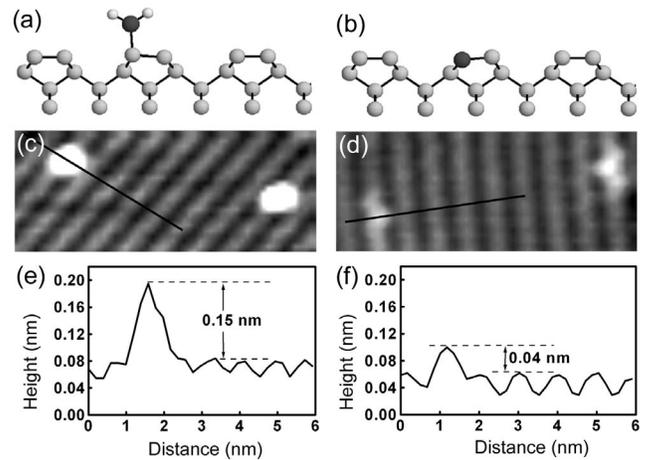

FIG. 2. Schematic diagrams and corresponding STM images at ~1.5 V of (a), (c) pairs of P containing molecules after adsorption onto the bare Si(100) surface and (b), (d) P atoms after incorporation into the surface as a result of annealing to ~ 530°C, with associated line profiles (e) and (f).

The remainder of the δ-doped sample was processed into a Van der Pauw geometry and Hall measurements were performed to determine the degree of electrical activation of the P dopants. Figure 3 (b) shows the 4.2 K sheet resistivity, $\rho_{xx}$ and Hall resistance, $\rho_{xy}$ of the δ-doped sample in magnetic fields up to 8 T. From the Hall slope we calculate the two-dimensional (2D) electron concentration to be $2.0 \pm 0.4 \times 10^{14}$ cm$^{-2}$. Comparing this with the P concentration of ~ $1.4 \times 10^{14}$ cm$^{-2}$ measured by SIMS and ~ $1.7 \times 10^{14}$ cm$^{-2}$ expected from saturation dosing and annealing, we conclude that the P atoms are completely electrically activated. This result highlights the importance of the critical anneal step and represents a significant improvement compared to results of other authors where complete electrical activation was not achieved.[9,11]

From the sheet resistivity we calculate the electron mobility of the 2D δ-doped layer to be ~$10^2$ cm$^2$/Vs. To date this is the highest reported mobility of a P δ-doped layer in Si. The strong peak in the sheet resistivity $\rho_{xx}$ at zero magnetic field is a clear signature of weak localization, which results from the confinement of electrons to the thin 2D δ-doped layer. This result confirms the 2D nature of the dopant distribution in agreement with the SIMS measurements.

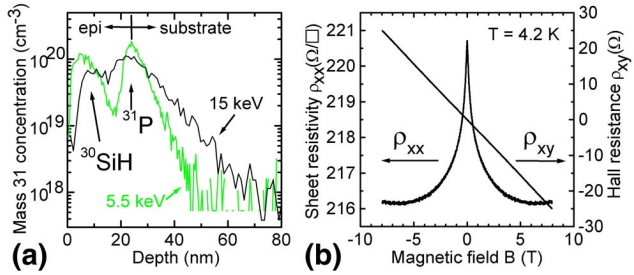

FIG. 3. (a) Mass 31 depth profile of the δ-doped layer determined by SIMS using 5.5 and 15 keV $Cs^+$ primary ion energies in an ATOMIKA system. (b) Sheet resistivity $\rho_{xx}$ and Hall resistance $\rho_{xy}$ of the δ-doped layer as a function of the magnetic field at T = 4.2 K.

Having demonstrated that silicon overgrowth temperatures as low as ~250 °C can limit dopant diffusion to less than only a few nm, we now consider if it is possible to further minimize diffusion so that STM patterned P arrays[6] can be maintained during high quality Si encapsulation. This may be achieved with more refined growth strategies, such as room temperature overgrowth with subsequent annealing. Initial results using room temperature growth and annealing are shown in Fig. 4. Figure 4 (a) shows an STM image of a Si surface that has been dosed with a low $PH_3$ dose, annealed at ~415 °C to incorporate the P atoms, encapsulated by 1 ML of Si grown at room temperature and then annealed to ~600 °C. We can clearly observe that the epitaxial surface is atomically flat and that the high anneal temperature allows P atoms to diffuse to the surface forming Si-P heterodimers (similar to those observed in Fig. 2 (d)). Here, we show for the first time direct evidence for the segregation/diffusion of individual P atoms to the Si surface to form Si-P heterodimers observed as the bright features in low bias STM images.[15] The P diffusion can be reduced by lowering the second anneal temperature to ~340 °C, as shown in Fig. 4 (b) where there are no apparent signs of the Si-P heterodimers, although it is harder to analyse this surface as it is not smooth. Investigations are currently underway to find a compromise between obtaining a flat surface for further high quality epitaxial growth and minimizing the diffusion/segregation of P atoms.

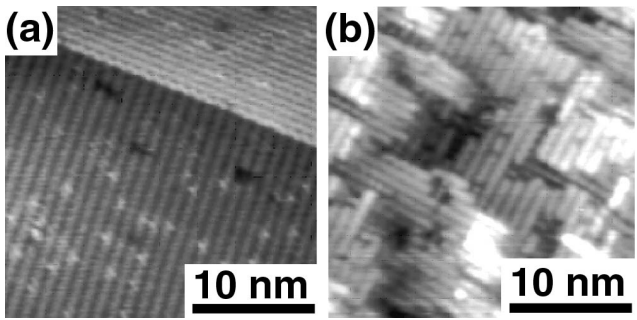

FIG. 4. STM images (bias –1.6 and –1.5 V, respectively) of a low $PH_3$ dosed layer, annealed and encapsulated by 1 ML of Si at room temperature and re-annealed at (a) 600 °C or (b) 340 °C.

In summary, we have demonstrated that phosphorus δ-doped layers can be encapsulated in silicon at growth temperatures as low as 250°C with minimal segregation of P dopants and complete electrical activation. The electron mobility of the δ-doped layer is ~$10^2$ $cm^2$/Vs with a strong weak localization peak around B = 0 demonstrating the 2D nature of the δ-layer. These results, along with future encapsulation strategies to further minimize dopant segregation, bode well for the fabrication of atomically precise dopant arrays in silicon where complete activation of the dopants is required. Future experiments will focus on epitaxial overgrowth of a well-defined array of single P atoms created by atomic lithography on a hydrogen terminated surface – the next significant step towards the realization of a Si based quantum computer.

The authors thank Dr. Gerhard Bilger (Institute of Physical Electronics, University of Stuttgart, Germany) for SIMS measurements. This work was supported by the Australian Research Council and the Army Research Office (ARO) under contract number DAAD19-01-1-0653.